\documentstyle[eqsecnum,aps]{revtex}
\begin{document}
\draft
\preprint{\vbox{\hbox{DOE/ER/40762-213}\hbox{UM PP\#01-026}}}
\title{Excited Heavy Baryons and Their Symmetries II: Effective Theory}
\author{Chi--Keung Chow, Thomas D.~Cohen and Boris A.~Gelman}
\address{Department of Physics, University of Maryland, College Park, MD 
20742-4111.}
\maketitle
\begin{abstract} 
We develop an effective theory for heavy baryons and their excited
states.  The approach is based on the contracted $O(8)$ symmetry recently
shown to emerge from QCD for these states in the combined large $N_c$ and 
heavy quark limits.  The effective theory is based on perturbations about this
limit; a  power counting scheme  is developed in which the 
small parameter is  $\lambda^{1/2}$ where $\lambda \sim 1/N_c \, , \, \Lambda
/m_Q$  (with $\Lambda$ being a typical strong interaction scale). We derive the
effective Hamiltonian for strong interactions at next-to-leading order. The 
next-to-leading order effective Hamiltonian depends on only two parameters
beyond the known masses of the nucleon and heavy meson. We also show that 
the effective operators for certain electroweak transitions can be obtained
with no unknown parameters at next-to-leading order.
\end{abstract}

\section{Introduction \label{I}}

Recently, it was pointed out that QCD when restricted to the space of heavy
baryons ({\it i.e.} states with baryon number one containing a single heavy 
--- b or c --- quark) has a contracted $U(4)$ \cite{hb0} or, more generally, 
$O(8)$ symmetry which emerges in the combined large $N_c$ (the number of 
colors) and heavy
quark limits \cite{hb1}. Of course, in the real world neither $N_c$ nor $m_Q$
(the generic heavy quark mass) is infinite. Nevertheless if they are large
enough, QCD will possess an approximate symmetry. Approximate symmetries can
provide significant phenomenological information about a system and have had a
long and distinguished history in strong interaction physics. Effective field
theory (EFT) is a powerful tool that can be used to extract this
information. In this article we develop an effective theory to
describe heavy baryons based on the approximate contracted $O(8)$ symmetry. 
This is important since this allows one to make testable predictions about the
phenomenology of heavy baryons. This paper serves the role of bridging
the formal developments of Ref.~\cite{hb1} with phenomenological descriptions
of heavy baryons. Our ultimate goal is to study the spectroscopy and 
electroweak decays of heavy baryon states which will be done in 
Ref.~\cite{hb3}.  

Before discussing how to implement the EFT approach in the context of heavy
baryons it is useful to recall how the EFT program works.  Effective field
theory is useful when there is a separation of scales between the low energy
phenomena and higher energy physics. The
short-distance physics can  greatly influence the long-distance physics.  
However, when working at low momentum one cannot resolve the details of what is
happening at short distances.  The idea of EFT is to describe the low-momentum
phenomena in a manner which is insensitive to the details of short-distance
physics.  Formally, one would begin with the full theory and
integrate out all of the short-distance degrees of freedom in a functional
integral.  This  yields an effective action to be used in a functional integral
over the low momentum (long-distance) degrees of freedom.  Clearly, the
effective action so obtained cannot be expressed as the integral over a local
Lagrangian.  However, all of the nonlocalities occur at the scale of the 
short-distance degrees of freedom which have been integrated.  Thus, to good
approximation the effective action can be written in terms of a local 
Lagrangian. The effects of the short-distance physics are now  contained in 
the coefficients in the effective Lagrangian.

Clearly such an approach is, at best, approximate. Fortunately, if the scale
separation between the long-distance scales and the short-distance scales is
large the approximation can be very good. Formally, one can develop a
systematic power-counting expansion where the  parameter is the ratio of the
typical long-distance scale to the typical short-distance scale.  As a
practical matter one must truncate the effective Lagrangian at some order.  The
power counting, however, ensures that the effects of the neglected terms are
formally down from the leading term by some power in the ratio of the scales. 
This, in
turn, ensures that the principal effects of the short-distance physics are
contained in a few coefficients in the effective Lagrangian. In this manner an
insensitivity to the details of the short-distance physics is achieved.  The
low momentum phenomenology depends on a few coefficients summarizing the
effects of the short-distance physics; the manner by which the short-distance
physics  produces these coefficients is irrelevant.

In principle, all of the  coefficients in the effective Lagrangian can be
calculated from the underlying physics.  Often, however, the underlying physics
is either unknown or is calculationally intractable.  EFT, nevertheless
provides a conceptual framework for making predictions about the system. The
key  insight is that one can treat the parameters in the effective Lagrangian
as phenomenological inputs fit from experiment.  Thus, providing there are more
pieces of independent data  than there are free parameters at a given order, 
one can make real, albeit approximate predictions.  Chiral perturbation
theory is a familiar paradigm for the EFT program. 

Let us now turn to the problem of heavy baryons. For simplicity we restrict 
our attention here to the problem of isoscalar heavy baryons --- the 
$\Lambda_b$
and $\Lambda_c$ (generically $\Lambda_Q$) and their excited states.  The
generalization to include non-isoscalars such as the $\Sigma_b$ or $\Sigma_c$
raises no new conceptual issues but complicates the analysis. The approach of
Ref.~\cite{hb1} begins with a restriction to a heavy baryon subspace of the
full QCD Hilbert space, {\it i.e.} we consider only the space of states with
baryon number one and one heavy quark. The physical picture that emerges from
the model-dependent analysis of Ref.~\cite{hb0,hb1} is that low-lying states 
of the heavy baryon in the combined large $N_c$ and heavy quark limits can be
represented as harmonic collective motion of the ``brown muck'' (the technical
term for the light degrees of freedom) against the heavy quark. An equivalent
picture in the combined limit is that of the model of 
Refs.~\cite{bst1,bst2,bst3,bst4,bst5,bst6,bst7,bst8,bst9} in
which the heavy baryons are treated as harmonic collective motions of an
ordinary baryon against a heavy meson.  The symmetry emerges quite simply from
this harmonic oscillator picture: the contracted $O(8)$ algebra is  the algebra
of all harmonic oscillator creation and annihilation operators and all
bilinears made from them.

The key to the analysis of Ref.~\cite{hb1} which enables this picture to 
emerge is a consistent power-counting scheme. Formally, since we are 
considering
the combined large $N_c$ and heavy quark limits, it is useful to consider
a single parameter which characterizes departures from this limit. Thus, we
introduce:
\begin{equation}
\lambda = \Lambda/m_Q \; , \; 1/N_c \, ,
\label{ct}
\end{equation}
where $\Lambda$ is a typical strong interaction scale and $m_Q$ is the mass of
the heavy quark.  The effective expansion parameter turns out not to be 
$\lambda$,
but $\lambda^{1/2}$ \cite{hb1}. The low-lying (collective) excitation energies
of the heavy baryons are of  order $\lambda^{1/2}$.  This should be contrasted
with excitation energies for internal excitation of the brown muck  and the
dissociation energy of the heavy baryon (into an ordinary baryon and a heavy
meson) which are each of order $\lambda^0$ \cite{hb1}. Thus, provided that
$\lambda^{1/2}$ can be considered to be small, a scale separation exists
between the low-lying collective states of the heavy baryons from all other
excitations.  This, in turn, implies that EFT methods can be
applied to this problem.  

We note at the outset, however, that in practice $\lambda^{1/2} \sim 3^{-1/2}$
is not a particularly small number so that the scale separation is not
particularly large.  This implies that one must work at relatively high order
in order to get relatively imprecise results. Thus for example, even working at
next-to-leading order one only expects relative accuracy nominally only of
order $\lambda$.  Clearly, the approach will be at best semi-quantitative. 
Moreover, the expansion will not converge for all observables.
On phenomenological grounds it is clear that at best the approach will be
useful for describing the ground state of the $\Lambda_Q$ and the doublet of 
the first excited orbital excitation. The second orbitally excited states 
(which have not been observed in either the $\Lambda_b$ or $\Lambda_c$ 
sectors) is either unbound or very near the threshold and clearly beyond the 
harmonic limit implicit in the effective field theory. The situation is 
reminiscent of the contracted $SU(2N_F)$ symmetries seen for light
baryons which emerge in the large $N_c$ limit \cite{SF1,SF2,SF3,SF4,SF5}. In
that case the large $N_c$ limit predicts an infinite tower of low-lying 
baryons. However, only the nucleon and delta for $N_f=2$ (or the octet and 
decuplet for $N_f=3$) are well described in the real world.  

The construction of the form of the effective Hamiltonian or Lagrangian for the
system is very straightforward.  Following the standard rules of effective
field theory, once the low-energy degrees of freedom are identified, one may
simply write down the most general Lagrangian built from them which is
consistent with the underlying symmetries and includes all terms up to a fixed
order in the power counting.  Similarly, one could add external sources to the
theory and impose power counting and thereby determine the from of currents. 
Of course, in order to compute observables one needs more than the form of the
effective Lagrangian and effective operators --- one needs values for
the coefficients.  These coefficients can either be obtained purely
phenomenologically or by matching with the underlying theory.  

In this paper we derive the form of the effective Hamiltonian at
next-to-leading order (NLO) directly from the structure of QCD and the 
power-counting rules .  We do this  formally by directly making a
change of variables to the collective degrees of freedom.  This has the benefit
of relating the coefficients in the effective theory directly to ground state
expectation values of known QCD operators. In this derivation the structure
of the Hilbert space plays an essential role. In particular, it is necessary to
show that in the combined limit, the Hilbert space can be written as a product
space of collective excitations of order $\lambda^{1/2}$ which may be 
interpreted as excitations of the brown muck moving coherently against the 
heavy quark, and non-collective excitation of order $\lambda^0$ which 
correspond to excitations of the brown muck itself. In addition, as we will 
show, the excitations of the two spaces are independent up to corrections of
higher order in $\lambda$.  

We also derive results about the effective theory of more phenomenological
interest.  One of the central achievements of this paper is to show that at NLO
the dynamics depends  on only two free parameters in the effective Hamiltonian
beyond known masses of the nucleon and heavy meson.  This is of
phenomenological significance in that the small value of the expansion
parameter suggests that even for quite crude calculations one should work at
least at NLO. Had there been a large number of free parameters at this order
it would have suggested that the scheme would be difficult to implement in
practice as data on heavy baryons becomes available. Another important result
of this paper is the demonstration that certain operators that are not 
contained in the Hamiltonian but which play a role in describing electroweak 
properties of heavy baryons are completely determined without the need for 
phenomenological coefficients (up to corrections of higher order in 
$\lambda^{1/2}$) from the known structure of QCD and the counting rules. This 
increases the predictive power of the effective theory by eliminating free 
parameters. 

As noted above, an expansion in $\lambda^{1/2}$ is unlikely to be rapidly
convergent. Thus it seems sensible to work at relatively high order in the
expansion. Here we have stopped at NLO for two reasons: one practical and one
theoretical. The practical reason is simply that beyond NLO the number of
parameters grows to the point where it is likely to exceed the number of 
observables which have any reasonable hope of being measured in the 
foreseeable future. The theoretical reason is that the change of
variables technique used to derive the effective theory in terms of known
operators in QCD is straightforward only up to NLO.  Beyond this order,
ambiguities in the definition of the heavy quark mass enter as do effects from
the coupling of the low-lowing states described by the effective theory with
higher energy excitations.  Thus, the treatment becomes more 
subtle and complex beyond this order.  Of course, this does not mean that
there is no valid effective theory beyond NLO.  However considerable
effort must be undertaken in going beyond this order and this effort
is simply not justified in view of the practical difficulties mentioned above.

This paper is  organized as follows. In Sec.~\ref{II} we follow 
Ref.~\cite{hb1} and briefly review
the derivation of the low energy collective variables for heavy baryons from
QCD with an emphasis on the power counting rules. In Sec.~\ref{III}, we show
that the structure of the Hilbert space as a product space emerges from
standard large $N_c$ analysis. The form of the effective Hamiltonian (up to
NLO) is derived from QCD in Sec.~\ref{IV}. Effective operators for
electroweak matrix elements are derived in Sec.~\ref{V}. Finally a brief 
summary is given in Sec.~\ref{VI}

\section{Dynamical Variables from Counting Rules and QCD \label{II}}

In this section we review the counting rules derived from QCD for heavy baryons
in the combined large $N_c$ and heavy quark limits. We will not reproduce the
derivations of Ref.~\cite{hb1} in detail. Rather, we will outline the basic 
strategy of the derivation along with the principal physical and mathematical
assumptions.  We also collect results which will be needed later in this paper.

The first physical point of significance is to restrict our attention to a 
limited part of the QCD Hilbert space --- the part with baryon number one
and heavy quark number one. Furthermore we will ultimately restrict our
attention to the lowest-lying states with these quantum numbers.  For
simplicity, in this section we will only consider a single species of heavy
quark at a time as the strong interaction does not couple different heavy
flavors. We will generalize to two heavy flavors when we consider electroweak
matrix elements in the following sections.  

QCD is a field theory and as such has an infinite number of degrees of freedom.
We can envision making a canonical transformation from the original quark and
gluon degrees of freedom to a new set of variables (at the quantum mechanical
level it becomes a unitary transformation).  The goal is to find such a
transformation which has the property that a certain set of the variables
(which we denote the collective variables), generate the low-lying collective
excited states when acting (repeatedly) on the ground state (the
$\Lambda_Q$).  Working explicitly with the collective variables one can
calculate directly the properties of the low-lying states. Because we
are working in the context of canonical transformations it is far more 
natural to work with the Hamiltonian formulation of the underlying quantum 
field theory (QCD) rather than the Lagrangian formulation. 

Unfortunately, there is no straightforward way to generate such a canonical 
transformation to collective variables exactly. However, it is easy to find a 
set of variables which behaves this way approximately.  That is, using the 
power-counting scheme of Eq.~(\ref{ct}) one can find collective variables 
which when acting on the ground state (repeatedly) generate the low-lying 
collective states plus a small component of higher excited states --- the 
amount of higher excited state admixed is suppressed by powers of 
$\lambda^{1/2}$.

One canonically conjugate pair of collective variables is the total momentum of
the system $\vec{P}$, the generator of translations (which is well defined in
QCD) and its conjugate variable $\vec{X}$, the generator of momentum boosts. 
This pair of collective variables, however, does not induce the internal
excitations of interest here.

One might also wish to introduce the momentum of the heavy quark as a 
collective variable.  The expression for $\vec{P}_Q$ appropriate in the heavy 
quark limit is 
\begin{equation} 
\vec{P}_Q \, = \, \int d^3 x \,  Q^{\dagger} (x) (-i \vec{D}) Q(x) \, ,
\label{PQ}
\end{equation} 
where $Q(x)$ is the heavy quark field, and $\vec{D}$ is the 
three-dimensional covariant derivative. In the heavy quark limit it
is apparent that the variable conjugate to $P_Q$ is $X_Q$ defined by 
\begin{equation}
\vec{X}_Q \, = \, \int d^3 x \, Q^{\dagger}(x) \,\vec{x}\, Q(x) \, .
\label{XQ}
\end{equation}
The definitions for  $\vec{P}_Q$ and $\vec{X}_Q$ in 
Eqs.~(\ref{PQ}), (\ref{XQ})
are clearly valid in the heavy quark limit for the subspace of states with a 
heavy quark number of unity; in this limit the heavy quark behaves as in 
nonrelativistic quantum mechanics. In the limit $\vec{P}_Q$ and $\vec{X}_Q$
correspond to the generators of translations and momentum boosts for the heavy
quark. Away from this limit the concept of boosting or translating the heavy 
quark separately from the full system is not strictly well defined and the 
$\vec{P}_Q$ and $\vec{X}_Q$ defined above need not be canonically 
conjugate. However, even away from the formal limit there will exist some 
canonically conjugate pair of operators which smoothly goes to the
variables defined in Eqs.~(\ref{PQ}), (\ref{XQ}) as the limit is
approached. Moreover, by standard heavy quark counting considerations this
canonically conjugate pair of operators will differ from those defined in
Eqs.~(\ref{PQ}), (\ref{XQ}) by an amount of order ${\cal
O}(\Lambda/m_Q) = {\cal O}(\lambda)$. Here we are interested in the theory up
to next-to-leading order, {\it i.e.} up to relative order $\lambda^{1/2}$. 
Accordingly, to the order at which we work we can take Eqs.~(\ref{PQ}), 
(\ref{XQ}) as unambiguous definitions.  

Unfortunately, the conjugate pairs $(\vec{X}, \vec{P})$ and 
$(\vec{X}_Q, \vec{P}_Q)$
are not independent.  Clearly $[\vec{P},\vec{X}_Q] \ne 0$ since $\vec{P}$ is a
generator of translations and translating the system necessarily translates 
the position of the heavy quark. Thus, we can not directly use  $(\vec{X}_Q,
\vec{P}_Q)$ as our collective variables to generate low-lying states.
However, from these two pairs we can generate two independent conjugate pairs.
In particular we can construct a linear combination of $\vec{P}$ and 
$\vec{P}_Q$, which we denote $\vec{p}$ and a linear combination of $\vec{X}$
and $\vec{X}_Q$, which we denote $\vec{x}$, such that $\vec{x}$ and $\vec{p}$ 
are conjugate to each other and commute with $\vec{P}$ and $\vec{X}$ up to 
corrections at next-to-next-to-leading order (NNLO). Intuitively $\vec{p}$ can
be thought of as the generator of translations of the brown muck relative to 
the heavy quark and $\vec{x}$ as its conjugate. As we will see below, the 
variables $(\vec{x}, \vec{p})$ do act as approximate collective variables in 
the sense outlined above.

The explicit construction of $\vec{p}$ and $\vec{x}$ is detailed in 
Ref.~\cite{hb1}. The key inputs to this construction are Poincare invariance,
heavy quark effective theory, and a decomposition of the QCD Hamiltonian
organized via power-counting. The decomposition of the QCD Hamiltonian is most
easily accomplished by thinking of the heavy baryon as a bound state of a 
heavy meson and a nucleon. Standard large $N_c$ QCD analysis allows us to 
identify the interaction energy as scaling as $N_c^0$, while the nucleon mass 
$m_N$ scales as $N_c$ \cite{LN1,LN2}. Standard heavy quark analysis implies 
that the interaction energy is of order $m_Q^0$ while the mass of the heavy 
meson, $m_H$, is given by
\begin{equation}
m_H \, = \, m_Q + {\cal O}(m_Q^0) \, . 
\label{mH}
\end{equation}
Thus, when constrained to the heavy baryon part of the Hilbert space, the QCD
Hamiltonian may be decomposed as:
\begin{equation} {\cal H} \, = \, m_H \, + \,
m_N \, + \, {\cal H}_{int} \, ,
\label{decomp}
\end{equation}
where ${\cal H}_{int}$ is of order $\lambda^0$. Poincare invariance implies 
that
\begin{equation}
 [X_j ,{\cal H}] \, = \, i  \, \frac{P_j}{\cal H} \, , 
\label{Poinc} 
\end{equation} 
while the standard heavy quark treatment implies that
\begin{equation}   
[{X_Q}_j, {\cal H}] \, = \, i \, \frac{{P_Q}_j}{\cal H} \, . 
\label{heavy}
\end{equation}
Using Eqs.~(\ref{decomp}), (\ref{Poinc}), (\ref{heavy}) along with 
Eq.(\ref{mH}) and
the fact that $m_N, m_H \sim \lambda$, the methods of Ref.~\cite{hb1} allows
one to deduce that
\begin{eqnarray} 
\vec{p} \, & = &\, \left( \frac{m_H}{m_H +m_N} \, \vec{P}  \, -  \, 
\vec{P}_Q \right ) \,\left  ( 1 \, + \, {\cal O}(\lambda) \right ) \, ,
\nonumber \\
\vec{x} \, & = &\, \left( \left ( 1 \, + \, \frac{m_H}{m_N} \right ) \, \
\left (\vec{X}  \, -  \vec{X}_Q \right) \right ) 
\,\left  ( 1 \, + \, {\cal O}(\lambda) \right ) \, . 
\label{xp}
\end{eqnarray}
The expressions for $\vec{x}$ and $\vec{p}$ differ from the equivalent
expressions in Ref.~\cite{hb1} in that we have exploited Eq.~(\ref{mH}) to 
express quantities in terms of $m_H$ which is directly physically accessible 
rather than $m_Q$ which is not. Throughout the remainder of this paper we will
often eliminate $m_Q$ in favor of $m_H$.

Thus, we have constructed two independent pairs of conjugate variables: 
$(\vec{x}, \vec{p})$ and $(\vec{X}, \vec{P})$.  These two pairs naturally 
arise in the Hamiltonian. However, as we will see in Sec.~\ref{V}, in 
describing electroweak operators these pairs are not so natural. Instead it is 
natural to use the positions and momenta of the heavy quarks 
$(\vec{X}_Q, \vec{P}_Q)$ and the corresponding independent set 
$(\vec{X}_\ell, \vec{P}_\ell)$ which correspond intuitively to the position 
and momenta of the light degrees of freedom. Formally they are defined 
as follows: 
\begin{eqnarray}
\vec{X}_\ell = \vec{x} + \vec{X}_Q \, , \nonumber \\
\vec{P}_\ell = \vec{P} - \vec{P}_Q \, .
\label{light}
\end{eqnarray}

In order to exploit the collective variables $\vec{x}$ and $\vec{p}$ to learn 
about the low-lying spectrum, it is useful to evaluate multiple commutators of
$\vec{x}$ and $\vec{p}$ with ${\cal H}$. Following Ref.~\cite{hb1} one can 
again use Eqs.~(\ref{mH}), (\ref{decomp}), (\ref{Poinc}), (\ref{heavy}) and 
the known $\lambda$ scaling of $m_N$ and $m_H$, {\it i.e.} 
$m_N, m_H \sim \lambda$, to obtain
\begin{equation}
[x_k, [x_j, {\cal H}]] \, = \, - \, \delta_{ij} / \mu \, \left ( 1 \, + \, 
{\cal O}(\lambda ) \right ) \, ,
\label{dcomx}
\end{equation}
where the reduced mass, $\mu$, is given by
\begin{equation} 
\mu \, = \, \frac{m_N  m_H}{m_N + m_H} \, .
\label{mu}
\end{equation}
More generally, it can be seen by the same methods that multiple commutators
of $\vec{x}$ with ${\cal H}$ are suppressed by multiple powers of $\lambda$.
In particular, if one takes $n$ commutators of components of $\vec{x}$ with 
${\cal H}$ with $n$  even, then the $n^{\rm th}$ commutator scales as 
$\lambda^{n-1}$. For example:
\begin{eqnarray}
[x_m,[x_l,[x_k, [x_j, {\cal H}]]]] \, & \sim & \lambda^3 \, ,
\nonumber \\ 
\,[x_o,[x_n,[x_m,[x_l,[x_k, [x_j, {\cal H}]]]]]] \, & \sim & \lambda^5  
\nonumber \\
& \cdots &
\label {xmultcom} 
\end{eqnarray}

In contrast, standard large $N_c$ counting implies that a displacement of the 
light degrees of freedom relative to the heavy degrees of freedom by a distance
of order $\lambda^0$ shifts the Hamiltonian by an amount of order $\lambda^0$:
\begin{equation} 
e^{-i \vec{p} \cdot \vec{y}} {\cal H} e^{i \vec{p} \cdot \vec{y}} \, - \, 
{\cal H} \, \sim \, {\cal O}(\lambda^0) \, ,
\label{displace}\end{equation}
where $\vec{y}$ is a c-number of order $\lambda^0$. Equation~(\ref{displace}) 
in turn implies that multiple commutators of $\vec{p}$ with $\cal H$ are 
generically of order unity:
\begin{eqnarray}
[p_k, [p_j, {\cal H}]] \, & \sim & \lambda^0 \, ,
\nonumber \\ 
\, [p_m,[p_l,[p_k, [p_j, {\cal H}]]]] \, & \sim & \lambda^0 \, ,
\nonumber \\ 
\,[p_o,[p_n,[p_m,[p_l,[p_k, [p_j, {\cal H}]]]]]] \, & \sim & \lambda^0  
\nonumber \\
& \cdots &
\label {pmultcom} 
\end{eqnarray}

The power-counting rules from QCD are essentially summarized in 
Eqs.~(\ref{xmultcom}) and (\ref{pmultcom}). By themselves, however, they are 
not sufficient to develop a fully systematic power-counting scheme for 
observables without additional phenomenological input. In particular, there 
are two possible scenarios in which the power-counting may be realized 
selfconsistently. Intuitively, the $O(8)$ or symmetric realization 
corresponds to the case where the effective potential between the heavy quark 
and the brown muck has a minimum at $\vec{x} =0$, while the 
$O(4) \,\otimes \,O(4)$ or symmetry broken realization
corresponds to the situation where the effective potential has a minimum at 
$\vec{x} \neq 0$. The  $O(4)\,\otimes \,O(4)$ realization is labeled symmetry 
broken since it breaks rotational symmetry in the $\lambda \rightarrow 0$ 
limit. As shown in Ref.~\cite{hb1} the $O(8)$ (symmetric) realization has the
generic property that
\begin{equation}
x \, \sim \, \lambda^{1/4} \, ,  \; \;
p \, \sim \, \lambda^{-1/4}
\label{generic}
\end{equation}
in the sense that typical matrix elements between low-lying states of 
polynomial operators which contain $x^{n_x}$ and $p^{n_p}$ scale
as $\lambda^{(n_x - n_p)/4}$.  In contrast, in the $O(4)\,\otimes \, O(4)$ 
realization,
\begin{equation}
x \, \sim \, \lambda^{0} \, ,  \; \; 
p \, \sim \, \lambda^{-1/4} \, .
\end{equation}
In this paper we will assume that the dynamics are such that the system is in 
the $O(8)$ (symmetric) realization.  

\section{Structure of the Hilbert Space \label{III}}

In this section we will show that in the combined large $N_c$ and heavy quark
limits, the Hilbert space can be written as a product space composed of
collective and intrinsic excitations, with the collective excitations having
energies of order $\lambda^{1/2}$ and intrinsic excitations are of
order unity.  The physical picture of these two types of excitations is quite
clear --- the collective excitations describe coherent motion of the brown muck
against the heavy quark while the intrinsic excitations are excitations of the
brown muck itself. 

The basic strategy in demonstrating that the Hilbert space is of this form
is to exploit the fact that our expansion parameter is valid in the pure
large $N_c$ limit where Witten's Hartree picture is well established 
\cite{LN2}.
The key point is that while nature is far closer to the heavy quark limit than
the large $N_c$ limit, our expansion is formulated in the combined large $N_c$
and heavy quark limits for $N_c \Lambda/m_Q$ arbitrary. Thus it applies whether
one takes the heavy quark limit first, the large $N_c$ limit first or takes the
limits simultaneously.  We will deduce the structure of the Hilbert space by
taking the large $N_c$ limit first and then argue that the result goes
over smoothly to the combined limit.  Ultimately, the reason that we can 
legitimately take the limits in any order stems from the fact that the dynamics
of the system are determined by the interaction Hamiltonian, 
${\cal H}_{\rm int} = {\cal  H} - m_H - m_N$.  Note that ${\cal H}_{\rm int}$
is non-singular in both the large $N_c$ and heavy quark limits, and therefore 
we can smoothly take either limit or the combined limit first.

Now consider taking the large $N_c$ limit prior to taking the heavy quark 
limit. In this case we can use Witten's Hartree picture of baryons. Each quark
propagates in the mean field of the remaining quarks. Low energy excitations 
are described by the excitations of a single quark in the background of the 
remaining quarks. In the present case we can divide these excitations into two
classes --- excitations of the heavy quark and excitations 
of one of the light quarks. Of course, for the light quark excitation it 
is necessary to anti-symmetrize the wave function as there are many light 
quarks. In contrast there is only one heavy quark. In the large $N_c$
limit these two classes of excitations are orthogonal  to each other so that 
the wave function may be written as a direct product:
\begin{equation}
|\psi \rangle_{\rm N_c \rightarrow \infty} \,= \, | h \rangle \otimes \ | 
\ell \rangle \, ,
\label{form}
\end{equation}
where $h$ ( $\ell$ ) represent the heavy (light) quark part of the Hartree 
wave function.  

The excitation energies for both types of excitations are of order $N_c^0$. 
However, the excitation energies of the heavy quark  are of order
$m_{Q}^{-1/2}$.  This is straightforward to see; the heavy quark sits in a 
potential well of depth and width independent of $m_Q$.  As one takes the
large $m_Q$ limit (after having taken the large $N_c$ limit), the heavy quark 
wave function becomes localized at the bottom of the potential well and thus
acts like an harmonic oscillator. One sees that the characteristic
excitation energies of the heavy and light types of motion scale with 
$\lambda$ according to
\begin{eqnarray}
\Delta E_h \, & \sim &\, N_c^0 m_Q^{-1/2} \sim \, \lambda^{1/2} \, ,
\nonumber \\
\Delta E_\ell \, & \sim  &\, N_c^0 m_Q^0 \sim \,  \lambda^0 \, .
\label{DeltaE}
\end{eqnarray}
As noted above, in this picture the excitations of the heavy quark 
and the light quarks are independent. The total excitation energy is simply 
the sum of the heavy excitation energy and the light excitation energy.

If we now consider the combined limit with the large $N_c$ and heavy quark
limits taken simultaneously the picture changes.  The Hartree approximation is
no longer valid.  However, the general counting rules remain --- they simply
have to be reinterpreted. Physically what happens in the combined limit is 
that instead of the heavy quark oscillating in the static field of the brown 
muck, as in the Hartree picture, they oscillate against each other. However,
the structure of the Hilbert space is not altered in its essence. The space
can still be broken up into a product space of independent classes of 
excitations as in Eq.~(\ref{form}), but now the two classes of excitations are 
not the heavy and light degrees of freedom as in the Hartree approximation. 
Rather, they are the collective excitations (between the brown muck and the 
heavy quark) and the intrinsic excitations (of the brown muck):
\begin{equation}
|\psi \rangle_{\lambda \rightarrow 0} \, = \, |{\rm C} \rangle \otimes \ 
|{\rm I} \rangle \, ,
\label{form2}
\end{equation}
where C ( I ) represent the collective (intrinsic) part. Similarly, the
excitation energies behave as in Eq.~(\ref{DeltaE}) but with the more 
general description of collective and intrinsic classes of excitations 
replacing the heavy and light classes of the Hartree approximation:  
\begin{eqnarray}
\Delta E_C \, & \sim & \,  \lambda^{1/2}  \, ,\nonumber \\
\Delta E_I \, & \sim & \, \lambda^0 \, .
\label{DeltaE2}\end{eqnarray}

We see that the structure of the Hilbert space is quite simple in the
$\lambda \rightarrow 0$ limit.  In order to understand the structure of the
effective theory, it is necessary to understand how corrections to this leading
behavior emerge.  In particular, we will need to quantify the mixing
between the collective and intrinsic excitations due to sub-leading terms in
$\lambda$.  Again, our strategy is to pass to the case where the large
$N_c$ limit is taken first and we have the Hartree approximation as a 
well-defined tool. Then use the smoothness of the two limits to pull back 
to the case of $\lambda \rightarrow 0$ keeping $N_c \Lambda / m_Q$ arbitrary.

It is useful to consider the way $1/N_c$ corrections emerge on top of the
Hartree picture as these corrections will ultimately become corrections in our
$\lambda$ counting. We begin by assuming that at finite, but large $N_c$, a 
Hartree picture is a good first approximation. We also make the standard large
$N_c$ assumption \cite{LN2} that corrections to the energy and to energy 
splittings are generically of relative order $1/N_c$. Moreover, these leading 
relative corrections are independent of $m_Q$ as the heavy quark only 
contributes 1 out of $N_c$ quarks. Thus, these relative corrections due to 
effects beyond the
Hartree approximation are necessarily of order $\lambda$ and do not contribute
at NLO (which is relative order $\lambda^{1/2}$). It is straightforward to
see that $1/N_c$ corrections to the energies imply that Hamiltonian matrix 
elements between two different states with different intrinsic states are
generically of order $N_c^{-1/2}$. This can be shown in two ways. First, 
off-diagonal matrix elements between different intrinsic states of
order $N_c^{-1/2}$ yield $N_c^{-1}$ corrections to the energies and we know 
corrections of this order exist. The second way is to construct an explicit 
wave function using a transition induced by the gluon exchange. The gluon 
exchange
contains two factors of the strong coupling and scales as $N_c^{-1}$; there is
a combinatoric factor of $N_c$ as the gluon can connect to any quark; and 
finally, there is a normalization factor of $N_c^{-1/2}$ because there are 
$N_c$ distinct terms in the
excited state wave function as any of the quarks may be excited.  Now we 
can pass from the large $N_c$ limit back to the combined $\lambda \rightarrow 
0$ limit (with $N_c \Lambda/m_Q$ fixed) by exploiting smoothness. We obtain
the following important result:
\begin{equation}
\langle {\rm  C , I } | {\cal H} | {\rm C , I' }\rangle \sim \lambda^{1/2}\, ,
\label{od}
\end{equation}
where the state $| {\rm C , I }\rangle$ is a product state
\begin{equation}
|{\rm  C , I} \rangle \, = \, | {\rm  C } \rangle \otimes | {\rm I } \rangle
\,.
\label{prod}
\end{equation}

\section{Construction of the Effective Hamiltonian \label{IV}}

In Sec.~\ref{II} we reviewed the power-counting rules for the 
collective variables $\vec{x}$ and $\vec{p}$ in the heavy baryon sector of the
QCD Hilbert space. As discussed in Sec.~\ref{I}, the general principle of 
effective field theory implies that the form of the effective theory is 
completely fixed by the symmetries of the theory and the power-counting rules.
Rather than simply invoking this general strategy, it is instructive to derive
the effective field theory at next-to-leading order directly from QCD through
the change in variables discussed in Sec.~\ref{II}. We do this because it is 
useful to verify that the proposed scaling rules are, indeed, selfconsistent;
the derivation gives significant insights into the underlying structure of the
theory; and finally, it is satisfying to verify explicitly the validity of the
general EFT philosophy in this case.

In the present context, where we have been implementing canonical
transformations to collective variables, it is simpler to work with an
effective Hamiltonian rather than an effective Lagrangian. From 
Eq.~(\ref{Poinc}) and the power-counting rules of Eqs.~(\ref{dcomx}), 
(\ref{xmultcom}), (\ref{pmultcom}) and (\ref{generic}) it is straightforward 
to deduce the form of the QCD Hamiltonian in the heavy baryon sector written 
in terms of the collective variables up to order $\lambda$ (NLO). It is given 
by
\begin{eqnarray}
{\cal H}_{\rm QCD } \,& = & \, m_H \,+ \, m_N \, +\,\frac{P^2}{2 (m_N + m_H)}
\,+ \,\frac{p^2}{2 \mu} + \,\hat{c}^{(0)} \, +\,\sum_j \hat{c}^{(1)}_j x_j \,
+ \, \sum_{j,k}  \hat{c}^{(2)}_{j,k} x_j x_k  \nonumber \\
& + & \, \sum_{j,k,l} \hat{c}^{(3)}_{j,k,l} x_i x_j x_l \, 
+\sum_{j,k,l,m} \hat{c}^{(4)}_{j,k,l,m} x_i x_j x_k x_l \, + 
\,{\cal O}(\lambda^{3/2} ) \, , 
\label{H}
\end{eqnarray} 
where $m_H$, $m_N$ $\sim \lambda^{-1}$, and $\hat{c}^{(a)} \sim \lambda^0$. 
From rotational invariance, the $\hat{c}^{(a)}$ are completely symmetric 
Cartesian tensor operators of rank $a$. Note that while terms up to $x^4$ 
contribute at this order, only terms up to $p^2$ contribute. The difference 
between these two is due to the essentially nonrelativistic nature of the 
kinematics. Formally, this difference is implicit in the different scaling 
relations for $x$ and $p$ in Eqs.~(\ref{dcomx}), (\ref{xmultcom}), 
(\ref{pmultcom}) and (\ref{generic}).

Since the form of Eq.~(\ref{H}) was deduced from commutation relations, 
one cannot immediately conclude that the $\hat{c}^{(a)}$ operators are 
c-numbers. Rather, at this stage in the derivation we can only deduce that 
they are operators which commute with all of the collective variables:
\begin{equation} 
[\vec{x},\hat{c}^{(a)}] \, = \, [\vec{X}, \hat{c}^{(a)}] \, = \, 
[\vec{p}, \hat{c}^{(a)}] \, = \,[\vec{P},\hat{c}^{(a)}] \, = \, 0 \, .
\end{equation}
Thus, the $\hat{c}^{(a)}$ tensors are, in general, operators which depend only
on the non-collective variables, {\it i.e.} the $\hat{c}^{(a)}$ are purely 
intrinsic operators. Note that to the extent that the coefficients 
$\hat{c}^{(a)}$ in the
Hamiltonian of Eq.~(\ref{H}) are operators as opposed to c-numbers, the
Hamiltonian is {\it not} strictly  an effective Hamiltonian. Rather it is the
full QCD Hamiltonian written in terms of our collective degrees of freedom. 
The non-collective degrees of freedom have not been integrated out; rather
their effects are contained in the non-collective operators $\hat{c}^{(a)}$.  

Non-collective operators  might be expected to influence the dynamics of
the collective degrees of freedom even though they all commute with the
collective degrees of freedom.  For example, consider the ground state matrix
element of $(\hat{c}^{(1)}_i x_i)^2$.  This can be expressed in terms of
off-diagonal matrix elements: 
\begin{equation} 
\langle 0 | (\hat{c}^{(1)}_i x_i) (\hat{c}^{(1)}_i x_i) | 0 \rangle \,  = 
\, \sum_{\alpha_C}\,\langle 0 | (\hat{c}^{(1)}_i x_i)| {\alpha_C}
\rangle \langle {\alpha_C} |  (\hat{c}^{(1)}_i x_i) | 0 \rangle \,
 + \, \sum_{\beta_{NC}}\,
\langle 0 | (\hat{c}^{(1)}_i x_i)| {\beta_{NC}} \rangle \langle {\beta_{NC}} |
(\hat{c}^{(1)}_i x_i)| 0 \rangle \, ,
\label{beta} 
\end{equation} 
where $| 0 \rangle$ is the ground state of the heavy baryon system and the
intermediate states are grouped into purely collective excitations 
$|\alpha_C \rangle$ (with excitation energies of order $\lambda^{1/2})$ 
which excite only the collective degrees of freedom leaving the brown muck 
unexcited, and non-collective states $| \beta_{NC} \rangle$
(with excitation energies of order $\lambda^{0}$ or greater) which involve an
intrinsic excitation of the brown muck. If the operators in the Hamiltonian
had been purely collective as in a true effective Hamiltonian (which by
construction has all non-collective degrees of freedom integrated out), then
the sum would not contain the non-collective states $| \beta_{NC} \rangle$. 
However, the non-collective operators $\hat{c}^{(a)}$ induce transitions 
to the non-collective states. By acting twice --- once to go into the 
non-collective space, and once to come back the collective space --- the 
non-collective operators can affect collective space matrix elements of 
operators contained in the Hamiltonian.

However, from Eq.~(\ref{od}) we know that terms which connect different
intrinsic states are suppressed by order $\lambda^{1/2}$. We can use this to
deduce that the contribution of non-collective intermediate states to matrix
elements of composite operators such as in Eq.~(\ref{beta}) will always be
suppressed by order $\lambda$ relative to the contributions of the collective
intermediated states and thus only contribute at NNLO. This in turn justifies
their omission at NLO. Let us see how this works for the decomposition in
Eq.~(\ref{beta}). From the scaling rules of Eq.~(\ref{generic}) it is
straightforward to see that the contribution from the collective intermediate
states goes as $\lambda^{1/2} $.  A typical term in the sum over states outside
the purely collective space is
\begin{equation} 
\left | \langle  0_{\rm C} ,
0_{\rm I}  |  c^{(1)}_i x_i  | {\rm C} ,  {\rm I} \rangle \right |^2 \, = \,
\left | \langle 0_{\rm C} | x_i  | {\rm C} \rangle \right |^2 \, \left |
\langle 0_{\rm I} | c^{(1)}_i  | {\rm I} \rangle \right |^2 \, \sim \,
\lambda^{1/2} \lambda \, ,
\label{sep}
\end{equation} 
where $| 0_{\rm C} \rangle $ and $| 0_{\rm I} \rangle $ represent the ground
states of the collective and intrinsic subspaces, respectively, while 
$| {\rm C} \rangle $ and $| {\rm I} \rangle $ 
represent excited states, and the notation
$|{\rm C}, {\rm I} \rangle$ represents a product state as in Eq.~(\ref{prod}) .
The equality in Eq.~(\ref{sep}) follows since $x_i$ only acts on the
collective space while $c^{(1)}_i$ only acts on the intrinsic space and the
scaling behavior follows from Eqs.~(\ref{generic}) and (\ref{od}) so that the
collective part of the expectation value scales as $\lambda^{1/2}$ while the
intrinsic part scales as $\lambda$. Thus, we see that the contribution of
non-collective states is down by a factor of $\lambda$ compared to the
contribution of the collective intermediate states.

The ${\cal O}(\lambda)$ suppression of the effects of non-collective
intermediate states relative to the collective ones is generic. Although it 
was explicitly demonstrated for the case of the composite operator in
Eq.~(\ref{beta}), it should be apparent that such a suppression occurs for all
products of operators contained in the Hamiltonian. When calculating
Hamiltonian matrix elements, the effects of the operators $\hat{c}^{(a)}$ in 
Eq.~(\ref{H}) inducing mixing with the non-collective states will always
contribute at relative order $\lambda$ ({\it i.e.} at NNLO) or less. Thus,
working at NLO one can simply replace the $\hat{c}^{(a)}$ operators by their
expectation values in the ground state (or more generally in the ground state
band --- the set of collective states built on the ground state of the 
intrinsic degrees of freedom). {\it Therefore, the operators $\hat{c}^{(a)}$ 
can be treated as c-numbers in the effective theory}.

When the coefficients $\hat{c}^{(a)}$ are treated as operators, rotational
invariance does not constrain the allowable forms in the Hamiltonian beyond
what is given in Eq.~(\ref{H}). For example, the term 
$\sum_i \hat{c}^{(1)}_i x_i$ is allowable despite the fact that $\vec{x}$ is an
$L=1$ operator; $\hat{c}^{(1)}$ could also be $L=1$ operator and couple to 
$\vec{x}$ 
leading to an $L=0$ operator in the Hamiltonian. Once the $\hat{c}^{(a)}$ 
operators are restricted to being c-numbers in the effective theory, rotational
invariance greatly restricts the allowable forms: c-numbers do not transform
under spatial rotations. Accordingly, the only surviving $\hat{c}^{(a)}$
coefficients are those which multiply rotational scalars. Thus, for example,
the $\hat{c}^{(1)}$ terms must vanish. Treating the $\hat{c}^{(a)}$ 
coefficients as c-numbers and imposing rotational symmetry yields the 
following form of the effective Hamiltonian up to ${\cal O}(\lambda)$:
\begin{equation}
\begin{array}{ccccccccccccccc}
 H_{\rm eff} & =  &  {\cal H} & =   & 
\frac{P^2}{2 (m_N + m_H)} & + & ( m_H \, + \, m_N )& \, + \, & {c}^{(0)} &\,+
\, & ( \frac{p^2}{2 \mu}  \, + \, \frac{1}{2} \kappa x^2 ) & \, + \, &
\frac{1}{4!} \alpha x^4 & \,  + \, & {\cal O}(\lambda^{3/2}) \, , \\ & & & &
\parallel & & \parallel & & \parallel & & \parallel & & \parallel & & \\ & & &
& {\cal H}_{\rm c.m.} & & {\cal H}_{\lambda^{-1}} & & {\cal H}_{\lambda^{0}} &
& {\cal H}_{\lambda^{1/2}} & & {\cal H}_{\lambda^{1}} & &
\end{array}
\label{Heff}
\end{equation}
where ${\cal H}_{c.m.}$ refers to the center of
mass motion of the entire system while ${\cal H}_{\lambda^n}$ refers to the
piece of the Hamiltonian whose leading contribution is of order $\lambda^n$. 
The coefficients $c^{(0)}$, $\kappa$ and $\alpha$ are of order $\lambda^0$
and may be expressed as ground-state expectation values of
$\hat{c}^{(0)}$, $\hat{c}^{(2)}$ and $\hat{c}^{(4)}$, respectively: 
\begin{eqnarray} 
c^{(0)} \,
&= & \langle 0 | \hat{c}^{(0)} | 0 \rangle \, , \nonumber \\ 
\kappa \, & = & 
\, \frac{2}{3} \, \sum_i \langle 0 | \hat{c}^{(2)}_{i,i} | 0 \rangle \, ,
\nonumber \\
\alpha \, & = &  \, \frac{24}{5} \, \sum_{i,j}\langle 0 |  
\hat{c}^{(4)}_{i,i,j,j}| 0 \rangle \, . 
\end{eqnarray}

A few comments about this derivation are in order here. In essence, we did not
have to explicitly integrate out the non-collective higher-lying degrees of
freedom. By making a change of variables to $\vec{x}$, $\vec{p}$, $\vec{X}$ 
and $\vec{P}$, we already have identified variables which were decoupled from 
the
intrinsic degrees of freedom to the order at which we are working. Imagine one
explicitly made this change of variables prior to formulating the problem in
terms of a functional integral. At low order the action could be written as
the sum of a collective action plus an intrinsic action without coupling
between them, and the only effect of integrating out the intrinsic degrees of
freedom would be to supply an overall constant multiplying the remaining
functional integral over the collective degrees of freedom. By making
this change of variables we make the effect of integrating out the intrinsic
degrees of freedom trivial --- at least up to NLO.  Only if we go beyond this
order do effects of the the coupling to the intrinsic degrees of freedom play
a role, and the effects of integrating out the intrinsic degrees of freedom 
become nontrivial.

From a phenomenological prospective, it is sufficient to start with
Eq.~(\ref{Heff}). The coefficients $c^{(0)}$, $\kappa$ and $\alpha$ could 
then be determined approximately from fits to experimental data. Given the 
fact that the masses of the nucleon and the heavy meson are known, the 
effective theory at NLO depends on three parameters --- $c^{(0)}$, $\kappa$ and
$\alpha$. The coefficient $c^{(0)}$ is determined from the difference in 
energy between the nucleon-heavy meson threshold and the lowest energy state. 
However, $c^{(0)}$ is simply an overall constant and plays no role in the 
dynamics; the dynamics of the problem depends on two parameters at NLO.
It is worth noting that at leading order the dynamics depends only on a single
parameter, $\kappa$. Thus, going from leading order to next-to-leading costs
only an additional parameter, $\alpha$.  

\section{Effective Operators \label{V}}

To make predictions of electroweak current matrix elements between heavy 
baryon states, one needs more than just the effective Hamiltonian. Effective 
operators are also required. For general composite operators expressed in terms
of the quark and glue degrees of freedom of QCD, the effective operators can 
be derived in a functional integral form by integrating out the intrinsic
degrees of freedom. The most straightforward way to formulate this is as an
integral over all QCD variables with $\delta$-functions put in to pick out
particular values of the collective variables. In practice, however, the
preceding description is rather formal as it is not generally possible to
perform this functional integration. At a phenomenological level, one can
simply write a form for the operator consistent with symmetries and
parameterized by some constants and then fit the constants from experimental
data. However, if the operators of interest can be expressed entirely in terms
of the collective degrees of freedom at the QCD level, the functional integral
approach becomes unnecessary (for the same reason it was unnecessary in the
derivation of the collective Hamiltonian). At NLO, one can immediately
express the effective operators in terms of the same collective degrees of
freedom.

To see how this works, consider some operator which at the QCD level can be
expressed entirely in terms of the collective operators.  We seek to find an
analogous operator for use in the effective theory whose matrix elements in the
effective theory reproduces the matrix elements of the QCD operator in the QCD
states (up to correction at NNLO). Let us start by considering the matrix
elements at the QCD level.  From the discussion of the structure of the Hilbert
space at the QCD level in Sec.~\ref{III}, it is apparent that a low-energy
({\it i.e.} with excitation energy of order $\lambda^{1/2}$) 
heavy baryon state in a $\lambda^{1/2}$ expansion at NLO can be written as:
\begin{equation}
| \psi \rangle \, = \,  |{\rm C} , 0_{\rm I} \rangle \, + \,
\lambda^{1/2} \sum_{{\rm C}',{\rm I} \ne 0_{\rm I}}
\, a_{{\rm C}', {\rm I}} \,| {\rm C}', {\rm I} \rangle \, ,
\label{psi}
\end{equation}
where $|{\rm C} , {\rm I} \rangle$ is defined as a product state as in
Eq.~(\ref{prod}), and $ a_{{\rm C}', {\rm I}}$ are constants of order unity.  

The sum in the second term of Eq.~(\ref{psi})  is restricted to contributions
with ${\rm I} \ne 0_{\rm I}$ due to the product structure of the state. 
If the sum were not restricted, then there would be contributions of the form
\begin{eqnarray}
& & |{\rm C} , 0_{\rm I} \rangle \, + \, \lambda^{1/2} \sum_{\rm C'}
\, a_{{\rm C}',  0_{\rm I}} \,| {\rm C}',0_{\rm I} \,\rangle =  \, 
| {\rm C} \rangle \otimes | 0_{\rm I} \rangle \, + \,
\lambda^{1/2} \sum_{\rm C'} \, a_{{\rm C}',  0_{\rm I}} \,| {\rm C}'\rangle
\otimes | 0_{\rm I} \,\rangle 
\nonumber \\ 
& & =  \, \left ( | {\rm C} \rangle \, + \, \lambda^{1/2} 
\sum_{\rm C'} \, a_{{\rm C}', 0_{\rm I}} \,| {\rm C}'\rangle \right ) \otimes |
0_{\rm I} \rangle \, =  | \tilde{\rm C}, 0_{\rm I} \rangle   \, , 
\label{dc}
\end{eqnarray}
where, $| \tilde{\rm C} \rangle = \left ( | {\rm C} \rangle \, + \, 
\lambda^{1/2} \sum_{\rm C'} a_{{\rm C}',  0_{\rm I}} \,| {\rm C}'\rangle 
\right )$ is simply a single state in the purely collective space. Thus, the 
contribution to the sum from Eq.~(\ref{dc}) is of the form of a single product
state. If there are contributions of the form as in Eq.~(\ref{dc}) we can 
replace $\rm C$ by $\tilde{\rm C}$ and thereby eliminate contributions 
with ${\rm I} = 0_{\rm I}$ from the sum. Doing this has the advantage of 
removing any possibility of double counting these contributions. 

Now let us evaluate the matrix element of a purely collective operator,
$\Theta_{\rm C}$, {\it i.e.} an operator constructed entirely of the 
collective operators $\vec{x}$, $\vec{p}$, $\vec{X}$ and $\vec{P}$. From the 
structure of the state in Eq.~(\ref{psi}), the matrix element between two 
low-energy heavy  baryon states (given at NLO) is given by
\begin{eqnarray}  
\langle \psi_1 | &\Theta_{\rm C}& | \psi_2 \rangle \,  = \, 
 \left (   \langle {\rm C_1} , 0_{\rm I_1} | \,  + \, 
\lambda^{1/2} \sum_{{\rm C_1}',{\rm I_1} \ne 0_{\rm I_1}}
 \, a_{{\rm C_1}', {\rm I_1}}^* \,\langle {\rm C_1}', {\rm I} 
| \right )  \Theta_{\rm C} \,  \left ( |{\rm C_2} , 0_{\rm I_2} \rangle \,  
+ \, \lambda^{1/2} \sum_{{\rm C_2}',{\rm I_2} \ne 0_{\rm I_2}}
 \, a_{{\rm C_2}', {\rm I_2}} \,| {\rm C_2}', {\rm I_2} 
\rangle \right )  
\nonumber \\ \nonumber \\
 & = & \, \langle  {\rm C_1} |  \Theta_{\rm C} | {\rm C_2} \rangle  \,
+\, \lambda \,  
\sum_{{\rm C_1}',  {\rm C_2}', {\rm I_1}\ne 0_{\rm I_1}, 
{\rm I_2}\ne 0_{\rm I_2}}
\delta_{{\rm I_1}, {\rm I_2}}
\, \, a_{{\rm C_1}', {\rm I_1}}^* \, \, a_{{\rm C_2}',{\rm I_2}} \, \,
\langle {\rm C_1}' | \Theta_{\rm C} | {\rm C_2}' \rangle \, = \,
\langle  {\rm C_1} |  \Theta_{\rm C} | {\rm C_2} \rangle \, 
\left ( 1 + {\cal O}(\lambda) \right ) \, .
\end{eqnarray}
Thus, up to NNLO corrections the QCD matrix element is simply the matrix 
element in the collective part of the Hilbert space. On the other hand, the 
effective Hamiltonian we use at this order with the operators $\hat{c}^{(a)}$ 
replaced by c-numbers is designed to correctly describe the collective part of
the Hilbert space at this order.  

The preceding argument shows, that in the case of purely collective operators 
the operators in the effective theory are related to the QCD operators in a 
trivial way up to corrections at NNLO. A purely collective operator at the 
QCD level is of the form 
$$ \Theta_{\rm C}^{\rm QCD}(\vec{x}^{\rm QCD}, \vec{p}^{\rm QCD},
\vec{X}^{\rm QCD},\vec{P}^{\rm QCD}) \, , $$
where the superscript QCD indicates that all of the operators are at the QCD 
level. The effective operator up to order $\lambda$
corrections is then obtained by keeping the same operator structure of 
$\Theta_{\rm C}^{\rm QCD}$, but replacing $\vec{x}^{\rm QCD}$, $\vec{p}^{\rm
QCD}$, $\vec{X}^{\rm QCD}$ and $\vec{P}^{\rm QCD}$ with the collective
operators  $\vec{x}$, $\vec{p}$, $\vec{X}$ and $\vec{P}$:
$$ \Theta^{\rm eff} \equiv
\Theta_{\rm C}^{\rm QCD}(\vec{x}, \vec{p}, \vec{X},\vec{P}) \, . $$
This is essentially a nonrenormalization theorem to this order.  The effects
of integrating out collective degrees of freedom only renormalize
the effective operators at NNLO.

In a similar way, all purely non-collective operators at the QCD level,
{\it i.e.} all operators which commute with all of the collective operators,
simply become c-numbers in the effective theory with a value equal to the
ground state expectation value of the analogous QCD operator.
These general principles can be used to deduce the form of several operators
in the effective theory which will prove useful in the calculation of
electroweak observables for heavy baryon states. 

Let us first consider the  operator which generates Lorentz boosts, $\vec{K}$.
The unitary operator for a Lorentz boost is $B_{\vec{v}} = e^{i \vec{v} \cdot
\vec{K} }$.  The boost operator is useful, for example, in calculating form 
factors which involve matrix elements of states with different momenta. 
One can use the boost operator to generate these states from a standard state 
in its rest frame. From Poincare invariance, the boost operator is given at 
the field theoretic level by 
\begin{equation} 
\vec{K} \, = \, \frac{1}{2} \, \left  (\vec{X} \,
{\cal H} + {\cal H} \, \vec{X} \right)\,\, =\,(m_N + m_H) \, \vec{X} \, + 
\, {\cal O}(\lambda) \, ,
\end{equation} 
where the second equality follows from
Eq.~(\ref{decomp}).  Thus, up to NLO at the QCD level $\vec{K} = (m_N +m_H)
\vec{X} $ and from the arguments given above, this can immediately be taken
over in the effective theory.  

Among the observables which can be studied using the combined expansion are
matrix elements of the electroweak current $J=\bar{Q}_j \Gamma Q_i$,
where subscripts $i$ and $j$ refer to the heavy quark flavors and $\Gamma$ is 
the Dirac structure of the left-hand current, 
$\Gamma=\gamma^\mu (1-\gamma^5)$. We will show here that the leading term in
the combined heavy quark and large $N_c$ expansion of the current $J$ can be 
expressed in terms of the collective variables 
$\vec{x}$, $\vec{p}$, $\vec{X}$, $\vec{P}$. Moreover, we will show that the 
lowest order corrections come at order $\lambda$ and not at 
$\lambda^{1/2}$, so that they contribute only at NNLO.

The combined expansion of the current $J$ can be done analogously to the pure 
heavy quark expansion in the Heavy Quark Effective Theory (HQET) 
\cite{HQ1,HQ2,HQ3,HQ4,HQ5,HQ6}. 
The idea is  to decompose the total heavy quark field into ``large'' and
``small'' components or ``light'' and ``heavy'' fields with the latter being 
suppressed by a factor of $1/m_Q$ relative to the ``large'' component. It is 
then possible 
to eliminate the ``small'' component in terms of the ``large'' component by 
means of a power series expansion. Near the combined limit considered here, 
the pure heavy quark expansion breaks down. It has been shown in 
Refs.~\cite{mQ1,mQ2,mQ3,mQ4} that
the next-to-leading order terms in the pure heavy quark expansion of the 
matrix elements of the current $J$ between the heavy baryon states contain 
coefficients proportional to $\bar{\Lambda}/m_Q$ where $\bar{\Lambda}=
m_{\Lambda_Q}-m_Q$ (with $m_{\Lambda_Q}$ being the total mass of a heavy 
baryon). In the HQET the constant $\bar{\Lambda}$ is treated as being much 
smaller than $m_Q$, so that the NLO operators are indeed suppressed. The 
situation is different, however, near the combined limit where $\bar{\Lambda}$
is of order $m_N$, so that $\bar{\Lambda} \sim \lambda^{-1}$ and 
$\bar{\Lambda}/m_Q \sim \lambda^0$. Hence, these corrections 
cannot be neglected near the combined limit. 

The appearance of large corrections near the combined limit is a result of the
way the total heavy quark field is separated into its ``large'' and ``small''
components. The definition of these components is necessarily different from
the analogous decomposition in HQET.  Near the combined limit, as in HQET, the
heavy quark is not far off-shell since its interaction with the brown muck is
of order $\lambda^0$ while its mass is of order $\lambda^{-1}$.  However, the 
space-time dependence of the heavy quark is much different near the combined 
limit. Near the pure heavy quark limit the brown muck mass is formally much 
less than the heavy quark mass. As a result, the space-time dependence of the 
heavy quark is essentially determined by the free particle Lagrangian. In HQET,
therefore, it is useful to redefine the heavy quark field by removing the 
phase factor of the solution of
the free Dirac equation, namely $e^{-im_Q v^\mu y_\mu}$. Near the combined 
limit the heavy quark remains essentially nonrelativistic. However, the 
space-time dependence of the heavy quark is much different from that of a 
free particle. The phase factor now should contain a contribution due to
the brown muck mass, which is formally of the same order as the heavy quark 
mass near the combined limit. This naturally leads to the following 
redefinition of the heavy quark field:
\begin{eqnarray}
&h_{Q}^{(v)}(y)  =  e^{-i(m_Q+m_N)\, v^\mu y_\mu }\, P_{+} \, Q(y) \, , & 
\nonumber \\
&H_{Q}^{(v)}(y)  =  e^{-i(m_Q+m_N)\, v^\mu y_\mu}\, P_{-} \, Q(y) \, , &
\label{ULCOMB}
\end{eqnarray}
where $v^\mu $ is the total 4-velocity of the heavy baryon, the operators 
$P_{\pm}=(1\pm\!\not\!\!v)/2$ in the rest frame of the heavy baryon project
out the upper and lower components of the heavy quark field, and $y$ here 
represents a space-time point. As in HQET, the fields $h_{Q}^{(v)}$ and 
$H_{Q}^{(v)}$ satisfy conditions $\not\!v h_{Q}^{(v)}=h_{Q}^{(v)}$ and 
$\not\!v H_{Q}^{(v)}=-H_{Q}^{(v)}$. Fields $h_{Q}^{(v)}$ and $H_{Q}^{(v)}$ 
are defined for each value of a heavy baryon
4-velocity $v^\mu$ as can be seen from Eq.~(\ref{ULCOMB}). The 
redefinition in Eq.~(\ref{ULCOMB}) has the effect of removing both the heavy
quark mass, $m_Q$, and the nucleon mass, $m_N$, from the dynamics, leaving a 
dynamical scale of order $\lambda^0$, {\it i.e.}, the scale of the interaction 
Hamiltonian near the combined limit which is 
${\cal H}_{int}={\cal H}-(m_Q+m_N)$. 

The heavy quark part of the QCD Lagrangian density in terms of the fields 
$h_{Q}^{(v)}$ and $H_{Q}^{(v)}$ has the form,
\begin{eqnarray}
{\cal L}_Q & = & \bar{Q}(i\!\not\!\! D -m_Q)Q  
 =  \bar{h}_{Q}^{(v)} \,( iv^\mu D_\mu)\,h_{Q}^{(v)} - 
\bar{h}_{Q}^{(v)} \,(iv^\mu D_\mu +2m_Q)\, H_{Q}^{(v)} + 
\bar{h}_{Q}^{(v)}\,( i{\!\not\!\! D}_\perp - 2m_Q) \,H_{Q}^{(v)} 
\nonumber \\
& + & \bar{H}_{Q}^{(v)} \,(i{\!\not\!\! D}_\perp)\, H_{Q}^{(v)} 
+ m_N \left (\bar{h}_{Q}^{(v)}\,h_{Q}^{(v)} - 
\bar{H}_{Q}^{(v)}\,H_{Q}^{(v)} + 
\bar{H}_{Q}^{(v)}h_{Q}^{(v)}-\bar{h}_{Q}^{(v)} H_{Q}^{(v)} \right ) \, , 
\label{LQCOMB}
\end{eqnarray}
where $D_{\perp}=D- (v^\nu D_\nu)v$ is the ``transverse part'' of the 
covariant derivative $D$.
The heavy quark Lagrangian in Eq.~(\ref{LQCOMB}) written in terms of the 
fields $h_{Q}^{(v)}$ and $H_{Q}^{(v)}$ differs from its analog in HQET due 
to the additional term proportional to $m_N$.  This $m_N$ term would 
naively suggest that the fields $h_{Q}^{(v)}$ and $H_{Q}^{(v)}$ are both heavy 
with masses $m_N$ and $m_Q+m_N$, respectively, which apparently prevents 
integrating out the ``heavy'' field $H_{Q}^{(v)}$. 

This problem is easily resolved if one considers the full Lagrangian density 
of the system:
\begin{equation}
{\cal L}={\cal L}_Q+{\cal L}_q+{\cal L}_{YM}=\bar{Q}(i\!\not\!\! D - m_Q)Q+ 
\sum_j \bar{q}_{j}(i\!\not\!\! D -m_j)q_j +{\cal L}_{YM} \, ,
\label{Lt}
\end{equation}
where the sum is over all light quarks, and ${\cal L}_{YM}$ is the 
Yang-Mills Lagrangian density. 
The total Lagrangian density in Eq.~(\ref{Lt}) should be re-expressed so as to
build the brown muck contribution into the heavy degrees of freedom. This is 
completely analogous to the Hamiltonian treatment in Sec.~\ref{III}. In that 
case it is done by adding and subtracting to the Hamiltonian a quantity which 
is an overall constant $m_N$ and regrouping terms according to their $\lambda$ 
counting scaling. In the Lagrangian formalism this can be accomplished in the 
following way using a Lorentz covariant operator $m_N \bar{Q}\!\not\!v Q$:
\begin{equation}
{\cal L}=({\cal L}_Q- m_N\bar{Q} \!\not\! v Q)+
({\cal L}_{q}+m_N \bar{Q}\!\not\!v Q 
+{\cal L}_{YM})={\cal L}_H+{\cal L}_\ell \, ,
\label{LtCOMB}
\end{equation}
where 
$${\cal L}_H \equiv {\cal L}_Q- m_N\bar{Q}\!\not\! v Q$$
and 
$${\cal L}_\ell \equiv {\cal L}_{q}+m_N \bar{Q}\!\not\! v Q +{\cal L}_{YM}\,.$$
In the rest frame of a heavy baryon operator, $\bar{Q} \!\not\!v Q$ is equal 
to the heavy quark density operator $Q^\dag Q$ and its spatial integral equals
to one in the Hilbert space of heavy baryons. Thus, in the Hamiltonian the 
additional term corresponds to removing from the light degrees of freedom an 
overall constant $m_N$. 

The additional operator proportional to the nucleon mass $m_N$ can easily be 
expressed in terms of the fields $h_Q^{(v)}$ and $H_Q^{(v)}$:
$$m_N\bar{Q}\!\not\!vQ=m_N \left ( \bar{h}_{Q}^{(v)}\,h_{Q}^{(v)}-
\bar{H}_{Q}^{(v)}\,H_{Q}^{(v)} +
\bar{H}_{Q}^{(v)}h_{Q}^{(v)}-\bar{h}_{Q}^{(v)} H_{Q}^{(v)} \right ).$$
This operator cancels the last term in Eq.~(\ref{LQCOMB}) so that the 
Lagrangian density ${\cal L}_H$ has the form:
\begin{equation}
{\cal L}_H=
\bar{h}_{Q}^{(v)} \,( iv^\mu D^\nu)\,h_{Q}^{(v)} - 
\bar{h}_{Q}^{(v)} \,(iv^\mu D^\nu +2m_Q)\, H_{Q}^{(v)} + 
\bar{h}_{Q}^{(v)}\,( i{\!\not\!\! D}_\perp - 2m_Q) \,H_{Q}^{(v)} +
\bar{H}_{Q}^{(v)} \,(i{\!\not\!\! D}_\perp)\, H_{Q}^{(v)}\, ,  
\label{LH}
\end{equation}
where the condition $\!\not\!vh_{Q}^{(v)}=h_{Q}^{(v)}$ is used.
Thus, the ``large'' component $h_{Q}^{(v)}$ describes a massless field while 
the ``small'' component $H_{Q}^{(v)}$ has the mass $2m_Q$. Using equations of 
motion the ``heavy'' field $H_{Q}^{(v)}$ can be expressed in terms of the 
``light'' field $h_{Q}^{(v)}$ as follows:
\begin{equation}
H_{Q}^{(v)}= {1 \over 2m_Q+iv^\mu D_\mu}i \,{\!\not\!\! D}_\perp h_{Q}^{(v)}
\, .
\label{HQ}
\end{equation}
The relation between fields $h_{Q}^{(v)}$ and $H_{Q}^{(v)}$ 
in Eq.~(\ref{HQ}) is identical in form to that in HQET between the ``large''
and ``small'' components of the heavy quark field $Q$. 

We can now expand the field 
$Q=e^{i(m_Q+m_N)v^\mu y_\mu} (h_{Q}^{(v)}+H_{Q}^{(v)})$, the Lagrangian 
density in Eq.~(\ref{LH}), and the current $J=\bar{Q}_j \Gamma Q_i$ in powers 
of $\lambda$. This can be done by eliminating fields $H_{Q}^{(v)}$ (using 
Eq.~(\ref{HQ})) and expanding the numerator in powers of $(iv^\mu D_\mu)/2m_Q$.
Thus, the combined expansion of the heavy quark field including terms up to 
order $\lambda$ is,
\begin{equation}
Q(y)=e^{-i(m_Q+m_N)v^\mu y_\mu} 
\left [ 1+ \left ({1-\!\not\!v \over 2} \right )
{i\!\not\!\!D \over 2m_Q} \right ] h_{Q}^{(v)}(y)\, + {\cal O}(\lambda^2).
\label{QCOMB}
\end{equation}
Using this expansion the Lagrangian density ${\cal L}_H$ including terms up 
to order $\lambda$ has the form:
\begin{equation}
{\cal L}_H = \bar{h}_{Q}^{(v)} \,(i v^\mu D_\mu) \,h_{Q}^{(v)} 
 + {1 \over 2m_Q} \bar{h}_{Q}^{(v)} \left [-(i v^\mu D_\mu )^2+(iD)^2 -
{1\over2}g_s \sigma_{\mu \nu}G^{\mu \nu} \right ]\, h_{Q}^{(v)} + 
{\cal O}(\lambda^2),
\label{LHCOMB}
\end{equation}  
where $g_s$ is a strong coupling constant and $G^{\mu\nu}=[D^\mu , D^\nu ]$ is 
the gluon field strength tensor.
In a similar way we can expand the current $J=\bar{Q}_j \Gamma Q_i$; again 
keeping terms up to $\lambda$ we get,
\begin{equation}
J(y=0)= \bar{h}_{Q_j}^{(v)}\Gamma h_{Q_i}^{(v)} + 
{1 \over 2m_{Q_i}} \bar{h}_{Q_j}^{(v)}\Gamma (i\!\not\!\! D)
h_{Q_i}^{(v)} + {1 \over 2m_{Q_j}} \bar{h}_{Q_i}^{(v)}\,
(-i\overleftarrow{\!\not\!\! D})\,
\Gamma h_{Q_i}^{(v)} +{\cal O}(\lambda^2) \, .
\label{JCOMB}
\end{equation}  
The expressions in Eqs.~(\ref{QCOMB}), (\ref{LHCOMB}), (\ref{JCOMB}) are 
identical in form to the analogous quantities in  HQET. However, there is an 
important difference: these expressions represent the combined 
expansion in powers of $\lambda$ and not the $1/m_Q$ expansion of HQET. 
In other words, by defining fields $h_{Q}^{(v)}$ and $H_{Q}^{(v)}$ 
appropriately we were able to resum implicitly all the $m_N/m_Q$ corrections 
in HQET.

The correction terms in the combined expansion of the electroweak current,
Eq.~(\ref{JCOMB}), come at order $\lambda$ and not at $\lambda^{1/2}$. 
It will be shown in Ref.~\cite{hb3} that these corrections
contribute to the matrix elements at order $\lambda$ as well, $i.e.$,
they do not induce any anomalously large constants which violate the combined
power counting of these operators. Thus, at NLO in our expansion only the 
leading order operator, $\bar{h}_{Q_j}^{(v)}\Gamma h_{Q_i}^{(v)}$, 
contributes to the matrix elements. In addition, it will be shown in 
Ref.~\cite{hb3} that in the combined limit at NLO there is only a single form 
factor which parameterizes the matrix elements of the current $J$ in the 
$\Lambda_Q$ sector. Accordingly, we can consider one Lorentz component of the 
current $J$, namely the $\mu =0$ component. Thus, to completely determine 
the matrix elements at NLO in the combined expansion, we need to know the form
of only one operator, namely  $h_{Q_j}^{(v)\dag} h_{Q_i}^{(v)}$. 

Let us first consider the flavor conserving operator
$h_{Q}^{(v)\dag} (\vec y)h_{Q}^{(v)}(\vec{y})$. As we are working near the 
heavy quark limit it is legitimate to expand the state in a fock-space
decomposition for the number of heavy quarks and antiquarks.  For a heavy
baryon state standard heavy quark analysis implies that the leading fock
component has one heavy quark; the first sub-leading fock component with two
quarks and one antiquark is suppressed by a factor of $\Lambda/m_Q \sim
\lambda$.  Thus at NLO in $\lambda^{1/2}$ we can neglect the sub-leading fock
component and treat the state as having a single heavy quark. This implies
that acting in our heavy baryon Hilbert space, the operator
$h_{Q}^{(v)\dag} (\vec y)h_{Q}^{(v)}(\vec{y})$ is 
$\delta^{3}(\vec{X}_Q-\vec{y})$. Since this is true at the QCD level at NLO it
immediately holds in the effective theory at that order.

In order to describe the flavor changing current we need to enlarge 
the Hilbert space of our effective theory to account for electroweak 
transitions between heavy baryons with 
heavy quarks of different flavors. The new Hilbert space is the direct 
product of the Hilbert spaces for baryons with a given flavor of a heavy 
quark described in Sec.~\ref{III}. Any state of the enlarged Hilbert space 
can be written as $n_f$-dimensional vector, where $n_f$ is a number of heavy
flavors. Each component of this vector describes a state with a given flavor. 
The effective Hamiltonian can be readily 
generalized to include baryons of various heavy quark flavors. It can be 
written as $n_f \times n_f$ diagonal matrix. For example, for two heavy 
flavors ($e.g.$ charm and bottom) the effective Hamiltonian including terms 
up to NLO in $\lambda^{1/2}$ has the form:
\begin{equation}
H_{\rm eff}=
\left (\begin{array}{cc}
{\cal H}_{Q_j} & 0 \\ 0 & {\cal H}_{Q_i}
\end{array}
\right) + {\cal O}(\lambda^{3/2}) \, ,
\label{matrH}
\end{equation}
where the diagonal entries ${\cal H}_{Q_j}$ and ${\cal H}_{Q_i}$ are the 
effective Hamiltonians as in Eq.~(\ref{Heff}). Analogously, QCD operators 
acting in the enlarged Hilbert space become $n_f \times n_f$ matrices. The 
flavor changing weak transitions can be described by off-diagonal matrix 
elements of an operator
$h_{Q_j}^{(v)\dag} (\vec y)h_{Q_i}^{(v)}(\vec{y})$, where the subscripts $i$ 
and $j$ can take integer values from $1$ to $n_f$. This operator can be 
written in terms of collective operators only. It follows from arguments 
analogous to the case of the flavor conserving electroweak operator ---
$h_{Q}^{(v)\dag} (\vec y)h_{Q}^{(v)}(\vec{y})$ --- that the flavor changing
operator acting on the enlarged Hilbert space has the form (for $n_f=2$):
\begin{equation}
h_{Q_j}^{(v)\dag} (\vec y)h_{Q_i}^{(v)}(\vec{y})=
\left (\begin{array}{cc}
\delta^{3}(\vec{X}_{Q_j}-\vec{y})\,\, &\,\, \delta^{3}(\vec{X}_{Q_j}-\vec{y}) 
\\ 
\delta^{3}(\vec{X}_{Q_i}-\vec{y})\,\, &\,\,\delta^{3}(\vec{X}_{Q_i}-\vec{y})
\end{array}
\right) \, + \, {\cal O}(\lambda) \, .
\label{WO}
\end{equation}
The diagonal elements of the operator in Eq.~(\ref{WO}) correspond to 
flavor conserving operators, while the off-diagonal elements
correspond to flavor changing operators. Their action is to 
instantaneously change the heavy quark flavor at the position where a heavy 
baryon interacts with the current. Since the operator in
Eq.~(\ref{WO}) is written in terms of the collective operators only, the 
corresponding effective operator has an identical form.

Finally, let us consider the operator defined by
\begin{equation} 
\vec{d}_\ell \, = \,  \frac{1}{N_c -1} \sum_{i = {\rm light}} \, \int\, d^3 x 
\, q_i^\dagger (\vec{x})\, \vec{x}\, q_i(\vec{x}) \, ,
\end{equation}
so that $\vec{d}_\ell$ represents the contribution to the baryon dipole moment
coming from light quarks.  In calculating electromagentic transitions
between heavy baryons it is useful to divide the electromagnetic current into
pieces coming from the heavy and the light quarks and to further divide the
light quark contribution into isovector and isoscalar pieces. The isoscalar
piece is simply the light-quark contribution to the baryon current. Thus, 
matrix elements of $\vec{d}_\ell$ must be calculated to describe electric 
dipole transitions of heavy baryons. 

At first sight, it might appear to be impossible to express the operator 
$\vec{d}_\ell$ in terms of collective operators. Although  $\vec{d}_\ell$ ---
the collective position operator for light quarks --- is intuitively something
like the position of the brown muck and clearly transforms in the same way as 
$\vec{X}_\ell$, at the operator level $\vec{d}_\ell$ is manifestly not 
identical to $\vec{X}_\ell$. Recall that by construction $m_N\vec{X}_\ell$ 
acts as a generator of boosts for the light degrees of freedom (up to order 
$\lambda$ corrections). On the other hand, the operator $\vec{d}_\ell$ acts 
only on the quarks and not on the gluons. Thus $m_N \vec{d}_\ell$ cannot boost
the gluons and therefore does not act as a  boost for the light degrees of 
freedom (which include both quarks and gluons). Thus, the operator 
$\vec{d}_\ell$ is clearly distinct from $\vec{X}_\ell$ and there is apparently
no simple way to express $\vec{d}_\ell$ entirely in terms of collective 
operators. 

Remarkably, however, one can show that 
\begin{equation}
\langle \psi' | \vec{d}_\ell | \psi \rangle \,  = \,
\langle \psi' | \vec{X}_\ell | \psi \rangle 
\label{dxequiv}
\end{equation}
for all $| \psi \rangle$, $| \psi' \rangle$ in the collective subspace defined
in Sec.~\ref{III}.  Thus, when one restricts consideration to the collective 
subspace which dominates the low energy physics, $\vec{d}_\ell$ is equivalent 
to $\vec{X}_\ell$.  

To see how this comes about, we start by considering the commutation relations
of $\vec{d}_\ell$ with $\vec{P}_\ell$, $\vec{X}_Q$ and $\vec{P_Q}$. To 
determine the commutator with $\vec{P}_\ell$ we exploit the fact that 
$\vec{P}_\ell$ is a generator of translations for the light degrees of freedom
so that ${\rm exp}(i \vec{P}_\ell \cdot \vec{\Delta}) \, q(\vec{x}) \, 
{\rm exp}(-i \vec{P}_\ell \cdot \vec{\Delta} )\,= \, q(\vec{x}-\vec{\Delta})$,
where $q$ is a generic light quark field. We can use this to deduce that
\begin{eqnarray}
& &{\rm exp}(i \vec{P}_\ell \cdot \vec{\Delta})\,\left (\vec{d}_\ell\right )\, 
{\rm exp}(-i \vec{P}_\ell \cdot  \vec{\Delta} ) \, \, =
\, \,  {\rm exp}(i\vec{P}_\ell \cdot \vec{\Delta} ) 
\left (  \frac{1}{N_c -1}\sum_{i = {\rm light}} \,
\int \, d^3 x \, q_i^\dagger (\vec{x}) \, \vec{x} \, q_i(\vec{x}) \right )\,
{\rm exp}(-i \vec{P}_\ell \cdot  \vec{\Delta} )\nonumber \\ \nonumber \\
& = & \, \left(  \frac{1}{N_c -1} \sum_{i = {\rm light}} \, \int\, d^3 x \, 
q_i^\dagger
(\vec{x} - \vec{\Delta} ) \, \vec{x} \, q_i(\vec{x} - \vec{\Delta}) \right ) 
\, \,  =  \, 
\,\left( \frac{1}{N_c -1} \sum_{i = {\rm light}} \, \int\, d^3 x' \, 
q_i^\dagger(\vec{x}' ) \, (\vec{x}' + \vec{\Delta}) \, q_i(\vec{x}') \right )
\nonumber \\ \nonumber \\
& = & \left( \frac{1}{N_c -1} \sum_{i = {\rm light}} \, \int\, d^3 x' 
\, q_i^\dagger(\vec{x}' ) \, \vec{x}'  \, q_i(\vec{x}') \, +  \, \vec{\Delta} 
\,  \frac{1}{N_c -1} \sum_{i = {\rm light}} \, \int\, d^3 x' 
\, q_i^\dagger(\vec{x}' )  \, q_i(\vec{x}') \right ) \, \, =  \, \, 
\vec{d}_\ell + \vec{\Delta} \, . 
\label{dtrans} 
\end{eqnarray}
The first equality in Eq.~(\ref{dtrans}) follows from the definition of 
$\vec{d}_\ell$, the second equality follows because $\vec{P}_\ell$ translates 
the light quark degrees of freedom, the third equality follows from a change 
of variables, and the final equality follows from the definition of 
$\vec{d}$ and the fact that 
$$ \sum_{i = {\rm light}} \, \int\, d^3 x' 
\, q_i^\dagger(\vec{x}' )  \, q_i(\vec{x}') $$
counts the net number of light quarks, which is $N_c -1$ in a heavy baryon
containing one heavy quark. Equation~(\ref{dtrans}) implies that 
\begin{equation}
[ {d_\ell}_j , {P_\ell}_k ] \, = \, i \delta_{jk} \, ,
\label{dcom1} 
\end{equation}
while the fact that from its definition $\vec{d}_\ell$ has no heavy quark 
operators implies that
\begin{equation}
[{d_\ell}_j , {X_Q}_j] \, = 0, \; \; [{d_\ell}_j , {P_Q}_k] = 0 \, .
\label{dcom2}
\end{equation}
Thus, the commutation relations of $\vec{d}_\ell$ with $\vec{P}_\ell$, 
$\vec{X}_Q$ and $\vec{P}_Q$ are identical to the commutation relations of 
$\vec{X}_\ell$ with $\vec{P}_\ell$, $\vec{X}_Q$ and $\vec{P}_Q$. In
Eqs.~(\ref{dcom1}), (\ref{dcom2}) indices $i, j = 1, 2, 3$ indicate Cartesian
components of $\vec{X}_\ell$, $\vec{P}_\ell$, $\vec{X}_Q$ and $\vec{P}_Q$.

Next let us express $\vec{d_\ell}$ in terms of the various collective and 
non-collective operators in the problem. Generically, one can represent an 
arbitrary operator ${\cal O}$ as a power series in our collective operators:
\begin{equation}
{\cal O} \, = \, \sum_N \hat{b}_N^{\cal O} \, \prod_{i=1,2,3} 
{P_\ell}_i^{N_{{P_\ell}_i}} \,
\prod_{i=1,2,3} {X_\ell}_i^{N_{{X_\ell}_i}} \,  
\prod_{i=1,2,3} {P_Q}_i^{N_{{P_Q}_i}} \,
\prod_{i=1,2,3} \,{X_Q}_i^{N_{{X_Q}_i}} \, 
\label{opser}
\end{equation}
with $N \, \equiv \, (N_{{P_\ell}_1}, N_{{P_\ell}_2}, N_{{P_\ell}_3},
N_{{X_\ell}_1}, N_{{X_\ell}_2}, N_{{X_\ell}_3},  
N_{{P_Q}_1}, N_{{P_Q}_2}, N_{{P_Q}_3},
N_{{X_Q}_1}, N_{{X_Q}_2}, N_{{X_Q}_3}) $ specifying a
particular term in the expansion, and $ i=1, 2, 3 $ indicating the three 
Cartesian directions. The coefficients in the expansion, $\hat{b}_N^{\cal O}$,
are generally non-collective operators which commute with 
$\vec{X}_\ell$, $\vec{P}_l$, $\vec{X}_Q$ and $\vec{P}_Q$. Expanding one 
component of $\vec{d}_\ell$ in this form gives
\begin{equation}
{d_\ell}_j \, = \, {X_\ell}_j \, + \,  \sum_{ N_{{P_\ell}_1}, \, 
N_{{P_\ell}_2}, \, 
N_{{P_\ell}_3} } \hat{b}^{{d_\ell}_j}_{N_{{P_\ell}_1}, \, 
N_{{P_\ell}_2}, \,N_{{P_\ell}_3} } \,
\prod_{i=1,2,3} {P_\ell}_i^{N_{{P_\ell}_i}} \, ,
\label{dlser}
\end{equation}
where the fact that the $\vec{d_\ell}$ commutes with  
$\vec{X}_Q$ or $\vec{P}_Q$ impose the restriction  that no nonzero powers of 
$\vec{X}_Q$ or $\vec{P}_Q$ can contribute to the series, while the commutation
relation of Eq.~(\ref{dcom1}) implies that the only contributing term 
containing any powers of $\vec{X}_\ell$ is the linear term and that the 
coefficient of this term is unity.

The known parity and time reversal properties of $\vec{d}_\ell$ and 
$\vec{P}_\ell$ fix the parity and time reversal properties 
of the non-collective coefficient operators,  
$\hat{b}^{{d_\ell}_j}_{N_{{P_\ell}_1}, \, N_{{P_\ell}_2}, \, 
N_{{P_\ell}_3} }$.  Since  $\vec{d}_\ell$ is time reversal even and parity odd
while $\vec{P}_\ell$ is time reversal odd and parity even, it follows that:
\begin{eqnarray}
T\left( \hat{b}^{{d_\ell}_j}_{N_{{P_\ell}_1}, \, N_{{P_\ell}_2}, \, 
N_{{P_\ell}_3} }  
\right ) \,&  = & \,
 (-1)^{N_{{P_\ell}_1} \, + \, N_{{P_\ell}_2} \, +  \, N_{{P_\ell}_3} } 
\nonumber \\ \nonumber \\
P\left( \hat{b}^{{d_\ell}_j}_{N_{{P_\ell}_1}, \, N_{{P_\ell}_2}, \, 
N_{{P_\ell}_3} }  
\right ) \, & = & \,
 (-1)^{N_{{P_\ell}_1} \, + \, N_{{P_\ell}_2} \, +  \, N_{{P_\ell}_3} \, + 1 } 
\label{PT} \end{eqnarray}
The significant point about Eq.~(\ref{PT}) is that every non-collective  
coefficient operator is odd under PT.  From the analysis of Sec.~\ref{III}, 
generic states can be written as superpositions of outer  product states 
containing collective and intrinsic parts. The collective subspace of the 
theory is the subspace where the intrinsic wave function is in its ground 
state.  Thus, an arbitrary state in the collective subspace can be written as 
$|\psi \rangle = |\psi_C \rangle \otimes | 0_I \rangle$ where $0_I$ indicates
the  ground state of the intrinsic system. Now consider the matrix element of 
a typical term in the series of Eq.~(\ref{dlser}) between two collective 
states:
\begin{equation}
\langle \psi' |  \hat{b}^{{d_\ell}_j}_{N_{{P_\ell}_1}, \, N_{{P_\ell}_2}, \,
N_{{P_\ell}_3} } \,
 \prod_{i=1,2,3} {P_\ell}_i^{N_{{P_\ell}_i}} | \psi \rangle \, = \, \langle 
\psi'_C|  
\prod_{i=1,2,3} {P_\ell}_i^{N_{{P_\ell}_i}} | \psi_C \rangle \langle 0_I |  
\hat{b}^{{d_\ell}_j}_{N_{{P_\ell}_1}, \, N_{{P_\ell}_2}, \,N_{{P_\ell}_3} } 
| 0_I \rangle \, , 
\label{van}
\end{equation}
where we have separated the matrix element into its collective and intrinsic 
parts. Clearly, the matrix element in Eq.~(\ref{van}) vanishes: the intrinsic
part of the of the matrix element is a diagonal matrix element of a PT odd 
operator in a state of good PT. Thus, within the collective subspace all terms
in the series of Eq.~(\ref{dlser}) vanish except the first term. This implies
that within this subspace all matrix elements of $\vec{d}_\ell$ are identical 
to matrix elements of $\vec{X}_\ell$ and Eq.~(\ref{dxequiv}) is established.

As written, the equivalence expressed in Eq.~(\ref{dxequiv}) only holds for 
states in the collective subspace, {\it i.e.}, states with the intrinsic 
subspace in its ground state. As discussed in Sec.~\ref{IV}, the 
non-collective operators in the Hamiltonian induce components with excited 
intrinsic states in the physical low-energy states.  This in turn spoils 
the formal  equivalence in  Eq.~(\ref{dxequiv}) for the physical states.  
Fortunately, however, it is straightforward to see that in the physical states,
the matrix elements of $\vec{d}_\ell$ differ from those of $\vec{X}_\ell$
by an amount of order $\lambda$ (or less) and hence may be neglected at 
next-to-leading order in $\lambda^{1/2}$. In the first place using reasoning 
analogous to that leading to Eq.~(\ref{od}), it is easy to see that the matrix
elements of the  
$\hat{b}^{{d_\ell}_j}_{N_{{P_\ell}_1}, \, N_{{P_\ell}_2}, \,N_{{P_\ell}_3}}$ 
between the
intrinsic ground state and its excited states is of order $\lambda^{1/2}$.  
On the other hand, the analysis of Sec.~\ref{IV} implies that admixtures of
excited intrinsic states in the physical states are suppressed by at least
order $\lambda^{1/2}$.  Combining these two facts one sees that the matrix 
elements of $\vec{d}_\ell$ and $\vec{X}_\ell$ in the physical states are
equivalent at NLO.

The equivalence at NLO between the matrix elements of  $\vec{d}_\ell$ and 
$\vec{X}_\ell$ may seem paradoxical. After all, as discussed above, the two
operators are clearly quite distinct at the QCD level. It would seem that the 
fact that the operators are different would imply that the matrix elements of
the two operators should be different in general. The resolution of the
paradox is quite simple: the matrix elements of the two operators
are different in general. They only coincide (up to NNLO corrections) for a
very special class of states --- the collective states with excitation
energies of order $\lambda^{1/2}$. For states with excitation energies of 
order unity, the matrix elements of the two operators can be quite different. 
Fortunately, we will be using our effective theory to study these low-lying
excitations only. Thus, for the purpose of the effective theory at NLO, it is
legitimate to replace $\vec{d}_\ell$ with $\vec{X}_\ell$.

\section{Summary \label{VI}}

In this paper we have derived an effective theory which will be used in 
Ref.~\cite{hb3} to study the spectroscopy and electroweak decays of the 
low-energy states of heavy baryons. The analysis is based largely on the 
formulation of the problem in Ref.~\cite{hb1}. A power-counting scheme is 
developed in $\lambda^{1/2}$, where $\lambda \sim 1 / N_c, \Lambda / m_Q$. 
The kinematical variables of the problem $\vec{X}$, $\vec{P}$, $\vec{x}$ and 
$\vec{p}$ are directly related to QCD operators which are well defined up to 
corrections at relative order $\lambda$. This formalism is based on standard 
large $N_c$ and heavy quark analysis.  

At the theoretical level, the analysis used in our derivation of the
effective theory  goes well beyond the analysis of Ref.~\cite{hb1} which was
largely kinematical and group-theoretical in nature. In our derivation, the
structure of the Hilbert space for heavy baryons plays a critical role.  
In particular, as the $\lambda \rightarrow 0$ limit is approached, the heavy 
baryon states can be written as products of collective states (with 
excitation energies of order $\lambda^{1/2}$) and intrinsic states (with 
excitation energies of order $\lambda^0$).  To deduce this form of the Hilbert
space we exploited the fact that our formulation as an expansion in $\lambda$ 
does not depend formally on whether the large $N_c$ limit is taken first, the 
heavy quark limit is taken first or they are taken simultaneously. By taking 
the large $N_c$ limit first we can use standard large $N_c$ reasoning to 
deduce the structure of the Hilbert space and then map back to the case where 
the limits are taken concurrently.  One essential point in this analysis is 
that matrix elements of operators which couple between different intrinsic 
states are
characteristically down by a factor of $\lambda^{1/2}$ compared to matrix
elements of operators with the same intrinsic state. Thus, the effect of
coupling to an excited intrinsic state and then back to the ground state is
typically of order $\lambda$ and hence can be neglected at NLO in the 
$\lambda^{1/2}$ expansion.

From the kinematics of the problem, the underlying heavy quark theory, 
Poincare invariance, and the structure of the Hilbert space, we showed that 
the effective Hamiltonian up to order $\lambda$ can be written in the form
of Eq.~(\ref{Heff}).  At order $\lambda^{1/2}$ one sees that the excitation 
spectrum  is fixed entirely by a single parameter $\kappa$, and at order 
$\lambda$ only one additional parameter --- $\alpha$ --- enters. This is
encouraging from a phenomenological perspective since we do not have a large 
number of coefficients to work with at NLO.

We also derived the certain effective operators at NLO. It turns out that the
operators discussed here are precisely those needed to compute interesting
electroweak transitions of  isoscalar heavy baryons at NLO. The key point in
the derivation is that all of these operators can be written in terms of the
collective operators (up to possible NNLO corrections). From this it was easy
to show that the effect of the admixture of components only comes in at NNLO 
so that at NLO one can simply replace the QCD level collective operator by an
effective operator of the same structure. One operator, $\vec{d}_\ell$, which 
is effectively the dipole moment of the light quark contribution to the baryon
density, was somewhat more subtle. Although, as an operator it cannot be
written directly in terms of the collective variables, we showed that its
matrix elements between low-lying states are identical to those of the 
collective operator $\vec{X}_\ell$.

The effective theory derived here is suitable for the study of isoscalar heavy
baryons at NLO in our $\lambda^{1/2}$ expansion. In Ref.~\cite{hb3} we will 
use the effective Hamiltonian of Sec.~\ref{IV} to study the spectroscopy and
electroweak decays of $\Lambda_c$ and $\Lambda_b$ baryons and their excited
states. In order to calculate the semi-leptonic matrix elements we will use 
relations between form-factors analogous to those obtained in HQET
\cite{HQ1,HQ2,HQ3,HQ4,HQ5,HQ6}. These matrix elements can then be determined 
using the effective operators discussed in Sec.~\ref{V}. We will also
calculate the radiative decay rates of the first excited states of $\Lambda_Q$
baryons.      

\bigskip 

\acknowledgments

We wish to acknowledge the support of the U.S.~Department of Energy under
Grant No. DE-FG02-93ER-40762.

\end{document}